# Association of Sex Hormone-Binding Globulin with Glycated Hemoglobin and Multidimensional Biomarker Structure in Reproductive-Age Women: An Analysis Based on NHANES 2021–2023

Ruilin Miao

*School of Statistics, Beijing Normal University, Zhuhai, China*

## Abstract

**Background.** Sex hormone-binding globulin (SHBG) is a hepatic glycoprotein that binds sex steroids and is associated with adiposity, insulin resistance, hepatic lipid metabolism, and type 2 diabetes risk. Nationally representative evidence on the shape of the continuous SHBG–HbA1c association in reproductive-age women, its age heterogeneity, and SHBG's position within a multidimensional biomarker system remains limited.

**Methods.** We analyzed NHANES August 2021—August 2023 data in nonpregnant women aged 20—44 years. Survey-weighted linear regression accounted for primary sampling units, strata, and the cycle-specific phlebotomy weights. SHBG was log-transformed and HbA1c was modeled continuously. Nonlinearity was evaluated with centered quadratic terms and natural cubic splines, and age modification was assessed for the linear slope and the full nonlinear curve. Sensitivity analyses considered adiposity, HOMA-IR, reproductive endocrine status, smoking, cotinine, sleep, a binary HbA1c outcome, and multiple imputation. Forty biomarkers were organized into hormone, biochemical, hematologic, and inflammation/lipid blocks and analyzed by multiple factor analysis (MFA), sparse PLS, LASSO with bootstrap stability selection, sJIVE, and nested cross-validation. Finite mixture regression (FMR) and mixture-of-experts (MoE) models were explored as secondary analyses of latent heterogeneity.

**Results.** The primary complete-case sample included 843 women. After adjustment for age, race/ethnicity, poverty-income ratio, and BMI, ln(SHBG) was inversely associated with HbA1c ($\beta$ =-0.1454, P=0.00083), corresponding to a 0.1008-percentage-point lower HbA1c per doubling of SHBG. Quadratic (P=0.00386) and spline (P=0.00236) tests supported nonlinearity; the cubic increment was not informative (P=0.826). The inverse association was strongest at lower SHBG concentrations and attenuated as SHBG increased. The ln(SHBG)×age interaction was significant (P=0.0169), whereas the joint test for age modification of the full nonlinear curve was not (P=0.0727). In a common sample, adjustment for BMI and waist circumference attenuated the SHBG coefficient by 43.7% and 48.1%, respectively. In the same fasting subsample, HOMA-IR adjustment attenuated the coefficient by approximately 55%. In multiblock analyses, the first three MFA dimensions explained 30.10% of total inertia. After SHBG was excluded from the latent-axis construction, ln(SHBG) remained associated with HbA1c after conditioning on the leading axes ($\beta$=-0.1062, P=0.00342). SHBG was selected in 98.8% of 500 LASSO bootstrap resamples at $\lambda$.min but only 38.8% at $\lambda$.1se. Nested cross-validation among 776 participants yielded out-of-fold $R^2$ values of 0.1298 (clinical model), 0.2136 (biomarker model), and 0.2392 (combined model). Outcome-guided sJIVE selected a joint rank of 0. Finite mixture and mixture-of-experts models failed prespecified minimum class-size and out-of-fold performance criteria relative to a single quadratic guard model.

**Conclusions.** In reproductive-age women, higher SHBG was associated with lower continuous HbA1c in a nonlinear pattern that was steeper at lower SHBG concentrations. The association overlapped substantially with adiposity and insulin-resistance-related variation but was not eliminated by covariate adjustment. Multiblock analyses placed SHBG at an endocrine–metabolic intersection and suggested additional conditional information beyond leading biomarker axes, while multidimensional biomarkers improved internal prediction beyond clinical variables. The cross-sectional design, limited fasting subsample, lack of external validation, and exploratory nature of multiblock and latent-class analyses preclude causal or clinical interpretation.

**Keywords:** Sex hormone-binding globulin; HbA1c; Reproductive-age women; NHANES; Complex survey; Multiple factor analysis; Multiblock biomarkers; Nested cross-validation

## 1. Introduction

Glycemic dysregulation is a major public health concern that typically develops over a prolonged subclinical phase before reaching diagnostic thresholds for diabetes. HbA1c provides an integrated measure of average glycemic exposure over approximately 8-12 weeks and can be assessed without fasting. The American Diabetes Association defines HbA1c of 5.7%—6.4% as prediabetes, formalizing recognition of this intermediate state [1]. Although HbA1c thresholds are clinically useful, analyzing HbA1c continuously retains information about intermediate metabolic variation that may be lost after dichotomization.

SHBG is synthesized primarily in the liver and transports sex steroids including testosterone and estradiol. Increasing evidence indicates that SHBG is linked not only to reproductive endocrinology but also to adiposity, hepatic lipid metabolism, inflammation, and insulin sensitivity [2-5]. Prospective studies have consistently associated higher SHBG with lower future diabetes risk, including among women [6,7], and longitudinal work in the SWAN cohort suggests that changes in SHBG across the menopause transition are themselves associated with incident diabetes [8].

Hormone concentrations vary with menstrual phase, reproductive status, pregnancy, and hormonal contraception. Young premenopausal women may also have little overt glycemic disease, making continuous metabolic phenotypes particularly informative. CARDIA data have shown that higher SHBG is associated with lower future metabolic syndrome risk in women aged 20—32 years at baseline [9], while a study of healthy postmenopausal women reported an inverse association between SHBG and HbA1c [10]. In contrast, a cross-sectional study of healthy cycling premenopausal women did not find a straightforward relationship between SHBG and HOMA-IR [11]. Most prior epidemiologic work has focused on postmenopausal or midlife women, clinical cohorts with polycystic ovary syndrome, or binary diabetes endpoints . Population-based evidence on the continuous SHBG–HbA1c association shape, age heterogeneity within the reproductive years, and the structural position of SHBG among multi-source biomarkers in a nationally representative sample of younger women remains limited.

We therefore analyzed publicly available data from NHANES 2021—2023 under a complex survey design to (i) quantify the conditional association between continuous SHBG and continuous HbA1c among non-pregnant women aged 20—44 years, (ii) examine nonlinearity and age-related effect modification, (iii) evaluate attenuation by adiposity and insulin resistance, and (iv) locate SHBG within a four-block biomarker space using unsupervised, supervised, and mixture modeling approaches.

# 2. Methods

## 2.1 Data Source and Study Population

NHANES is a continuous, nationally representative survey of the non-institutionalized U.S. civilian population conducted by the National Center for Health Statistics. The survey employs a complex, stratified, multistage probability sampling design and releases public-use data files that include sampling weights, strata, and primary sampling units (PSUs) [12]. We used the 2021—2023 cycle files linked by the unique participant identifier SEQN.

The target population was women aged 20—44 years with negative pregnancy testing at examination and WTPH2YR>0. The primary analysis required valid SHBG, HbA1c, age, race/ethnicity, PIR, and BMI and included 843 complete cases. A fasting analysis required a valid fasting weight and fasting glucose and insulin measurements. The multiblock analysis included 899 participants with complete data on 40 core biomarkers and 898 with additional HbA1c measurements. Nested prediction used 776 complete cases across clinical covariates, biomarkers, and HbA1c.

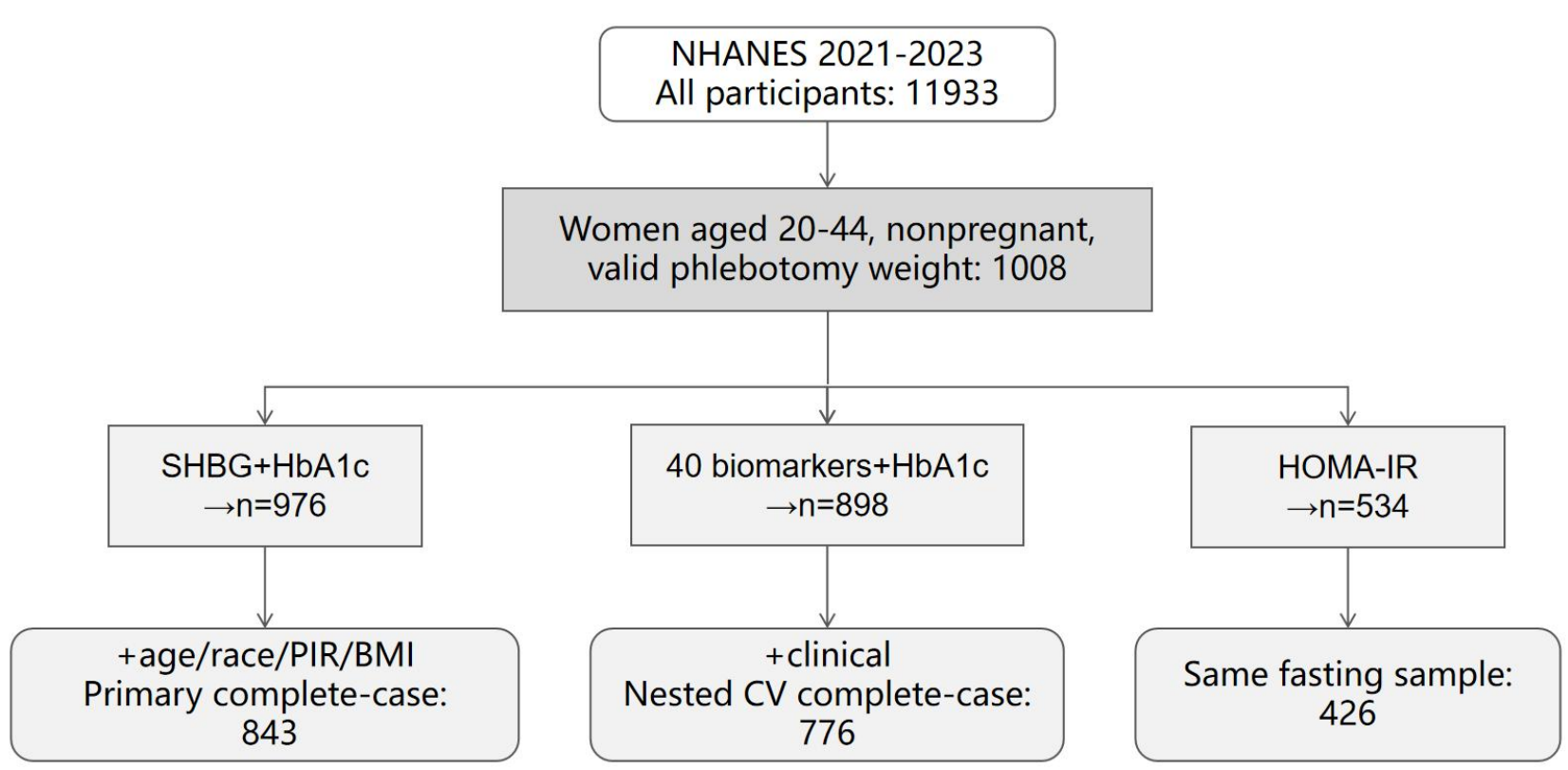

Figure 1. Study population and analytic sample construction.

## 2.2 Exposure, Outcome, and Covariates

Serum SHBG was the primary exposure and was natural-log transformed. HbA1c (%) was the primary continuous outcome. The primary adjustment set included age, race/ethnicity, PIR, and BMI. Additional sensitivity variables included waist circumference, HOMA-IR, progesterone, menstrual status, smoking, serum cotinine, sedentary time, and sleep duration. HOMA-IR was calculated as fasting glucose (mg/dL) × fasting insulin (μU/mL)/405 [13]. A binary outcome was defined as HbA1c≥5.7%.

## 2.3 Survey-weighted association models

We constructed survey design objects using SDMVPSU, SDMVSTRA, and WTPH2YR for the main blood-analyte analyses and WTSAF2YR for fasting analyses. Variances were estimated by Taylor linearization. The primary model was $HbA1c = \beta 0 + \beta 1 \ln(SHBG) + \gamma^T Z + \varepsilon$. Coefficients were reported per unit increase in ln(SHBG) and per doubling of SHBG, calculated as $\beta 1 \times \ln(2)$.

Nonlinearity was assessed by adding a centered quadratic term, testing the incremental cubic term, and fitting a four-degree-of-freedom natural cubic spline. Local slopes were derived from the quadratic model. Age modification was examined using an ln(SHBG) × age interaction and, separately, a joint test of age interactions with both the linear and quadratic SHBG terms. BMI-stratified estimates were treated as descriptive because the formal interaction was not significant.

## 2.4 Sensitivity analyses

Adiposity sensitivity used a common sample of 814 participants to compare a model without an adiposity measure with models adding BMI or waist circumference. HOMA-IR analyses were performed in a fasting subset; because of the smaller sample and cross-sectional design, HOMA-IR was treated as a metabolic sensitivity variable rather than a causal mediator. Additional analyses restricted progesterone levels, menstruation within the previous 12 months, smoking, cotinine, sedentary behavior, and sleep duration; assessed a binary HbA1c outcome; and repeated the primary model after multiple imputation. Twenty imputed datasets were generated using predictive mean matching and combined with Rubin's rules.

## 2.5 Multiblock biomarker analysis

Forty biomarkers were organized a priori into four blocks: hormones (12), biochemistry (18), complete blood counts (7), and inflammation/lipids (3). Variables were standardized within blocks and transformed as needed. Weighted blockwise PCA described within-block structure; parallel analysis used 500 random replicates. MFA was then used to integrate the blocks while balancing block contributions [14].

To test whether SHBG carried information beyond the main latent biomarker structure without mathematical coupling, SHBG was removed and MFA was reconstructed from the remaining 39 biomarkers; the leading axes were then entered as covariates with ln(SHBG). sPLS provided supervised dimension reduction [15], while LASSO [16] with 500 bootstrap resamples provided variable-selection stability. Prediction was evaluated using nested cross-validation: 10 outer folds and 5 inner folds, with all preprocessing and tuning confined to the training folds. Nested CV was used because conventional reuse of cross-validation for tuning and evaluation can yield optimistic error estimates [17].

sJIVE was used to compare joint versus block-specific outcome-relevant structure [18]. Finite mixture regression and age/BMI-gated MoE models were treated as exploratory latent-heterogeneity analyses. A single quadratic model served as a guard. Mixture models were required to have a minimum class proportion of 10% and were evaluated using strict OOF prediction and bootstrap classification stability; test-set outcomes were not used to assign latent classes.

## 3. Results

### 3.1 Primary Conditional Association

After sequential application of sex, age, pregnancy, and weight criteria, 1 008 women formed the blood-analysis target population; 843 complete cases entered the primary association analysis. Weighted mean age was 32.5 years (SE 0.37), BMI 29.5 kg/m² (SE 0.51), and HbA1c 5.35% (SE 0.04). The weighted prevalence of HbA1c ≥ 5.7% was 14.95% (SE 1.91). Weighted SHBG percentiles were 34.3 (P25), 50.5 (P50), and 77.7 nmol/L (P75).

Table 1. Survey-weighted main model of ln(SHBG) and HbA1c

| Model | β ln(SHBG) | P | HbA1c difference per doubling |
|---|---|---|---|
| Crude | -0.2585 | <0.001 | -0.1792 |
| +age | -0.2622 | <0.001 | -0.1818 |
| + race/ethnicity + PIR | -0.2328 | <0.001 | -0.1614 |
| +BMI | -0.1454 | 0.00083 | -0.1008 |

Across all adjustment levels, the SHBG association remained negative. In the fully adjusted model, each doubling of SHBG corresponded to an average HbA1c difference of -0.1008 percentage points. Approximately 37.5% of the coefficient magnitude observed in the model without BMI was attenuated upon BMI inclusion.

### 3.2 Nonlinearity

The quadratic term was significant (P=0.00386), the cubic increment was not (P=0.826), and the natural spline was significant overall (P=0.00236). Local slopes were -0.441, -0.250, -0.096, -0.027, +0.052, and +0.193 at SHBG values of 20, 40, 70, 90, 120, and 200 nmol/L, respectively. The data therefore support a steep inverse association at low SHBG with progressive attenuation at higher levels.

The curve shows the nonlinear pattern estimated in the fully adjusted survey-weighted model.

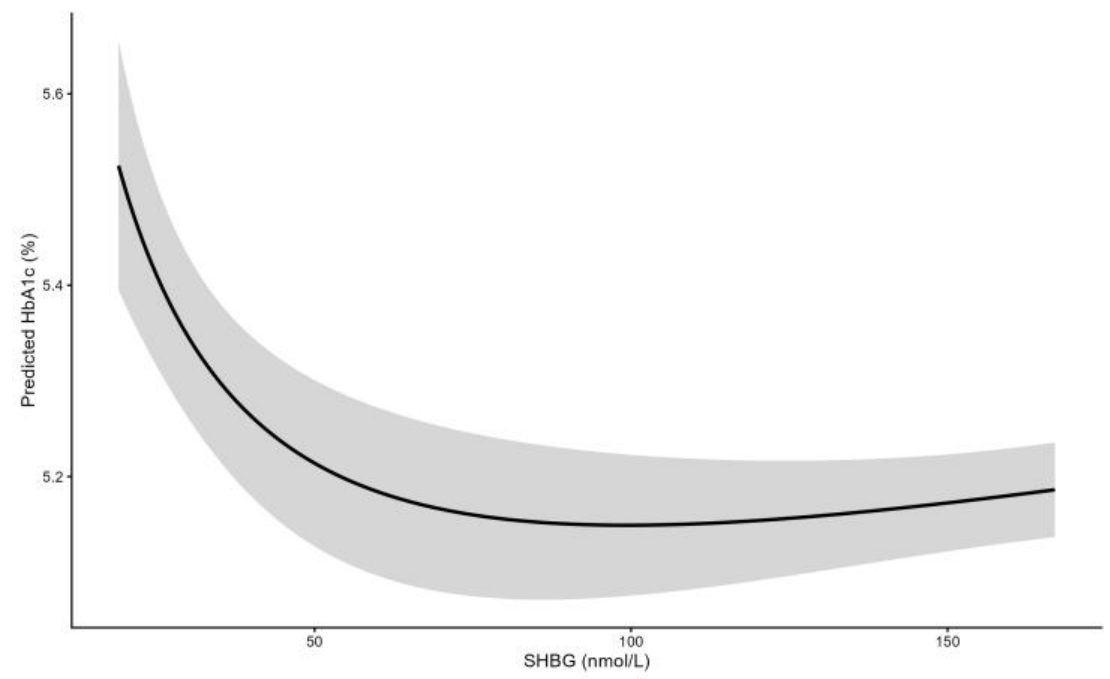

Figure 2. Covariate-adjusted SHBG–HbA1c association.

### 3.3 Age Heterogeneity

The ln(SHBG) × age interaction was statistically significant (P=0.0169). The conditional linear slope shifted from approximately +0.036 at age 20 to -0.319 at age 44. However, the joint test of age interactions with the linear and quadratic SHBG terms was not significant (P=0.0727), so the evidence supports age-related heterogeneity of the linear slope more strongly than modification of the entire nonlinear curve.

Age-stratified estimates were -0.0256 (20—29 years), -0.1276 (30—39 years), and -0.4123 (40—44 years).

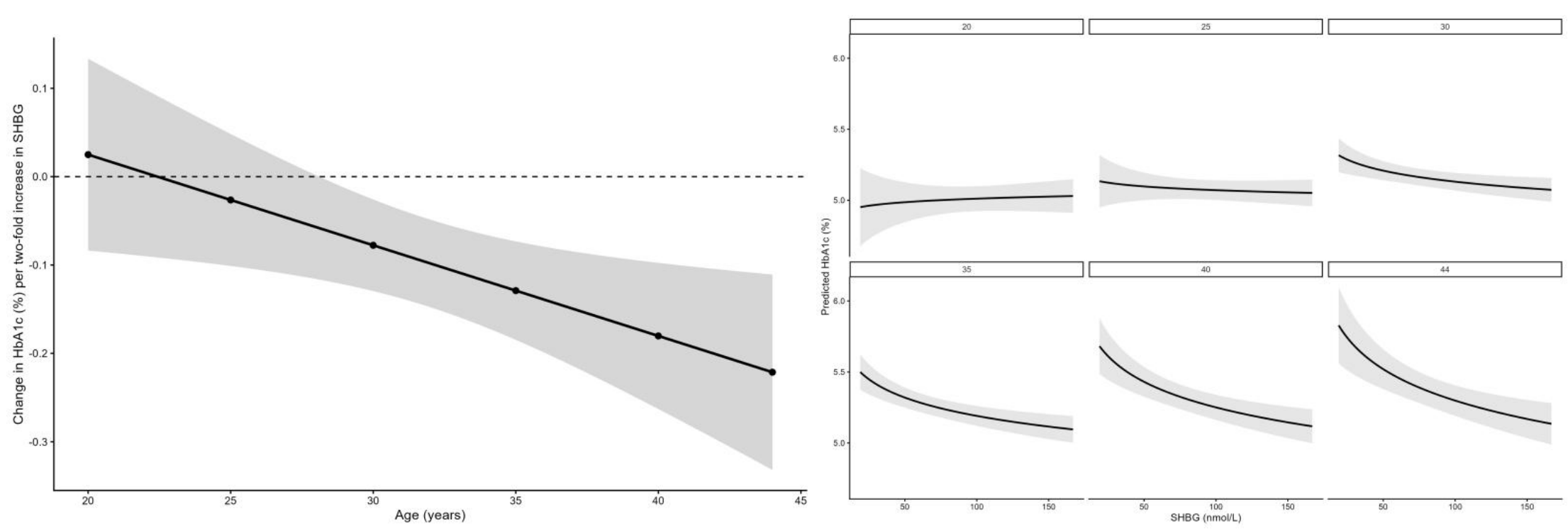

Figure 3. Estimated linear SHBG–HbA1c slope across age, adjusted for race/ethnicity, PIR, and BMI.

### 3.4 Adiposity, Insulin Resistance, and Sensitivity Analyses

On the common sample of 814 women, addition of BMI attenuated the SHBG coefficient by 43.7% and addition of waist circumference by 48.1%; both adjusted associations remained statistically significant.

In the fasting subsample, ln(SHBG) was inversely associated with log(HOMA-IR); addition of log(HOMA-IR) attenuated the SHBG–HbA1c coefficient by approximately 55%. Restrictions based on progesterone level, recent menstruation, or both left the inverse association directionally intact and statistically significant.

Addition of smoking status, cotinine, sedentary time, or sleep duration produced coefficients ranging from -0.123 to -0.155. Binary logistic models for HbA1c ≥ 5.7% yielded an odds ratio of 0.41 per unit ln(SHBG) (approximately 0.54 per doubling). Twenty-fold multiple imputation produced a combined estimate of -0.1516, closely consistent with the complete-case result.

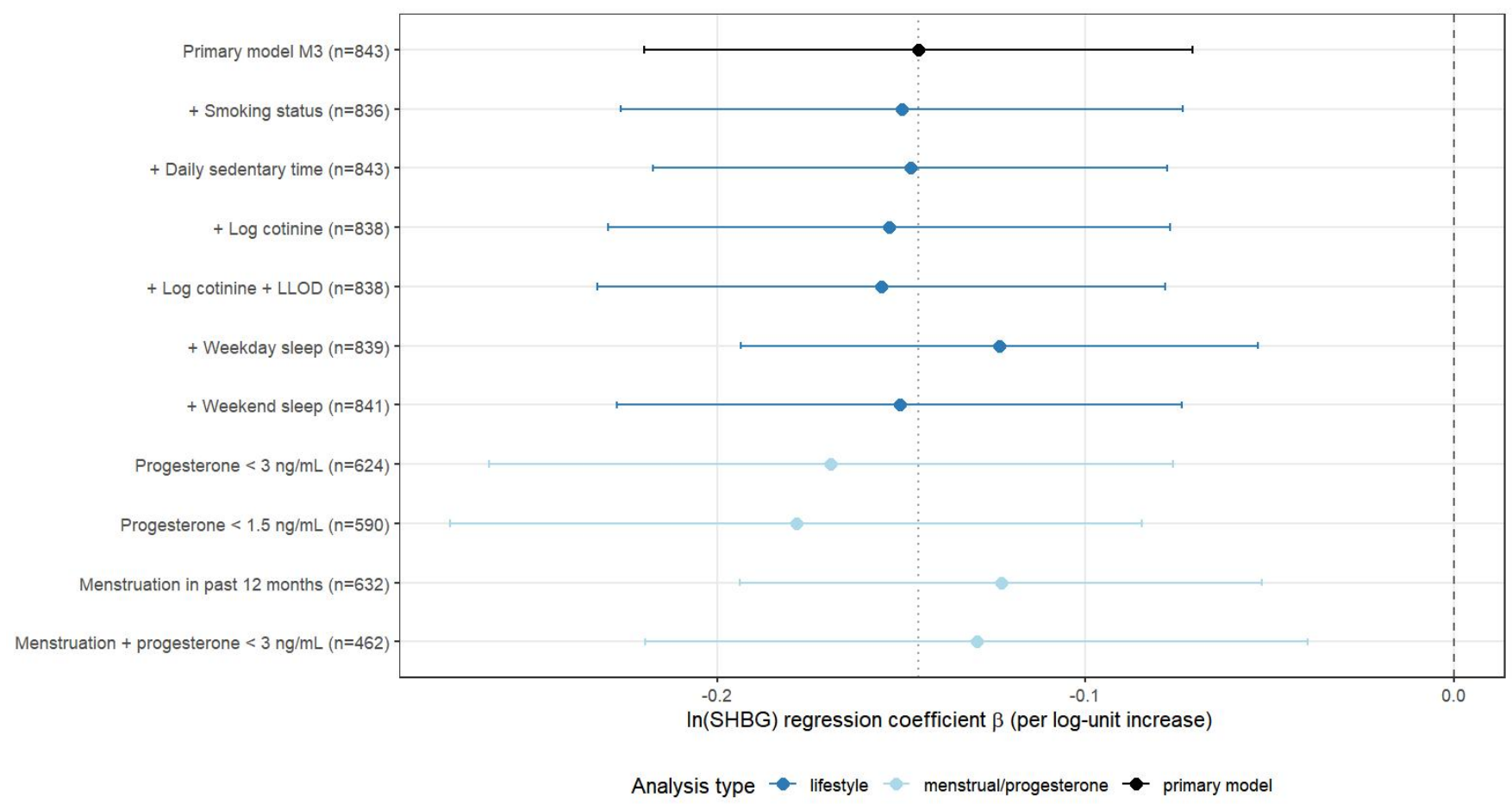

Figure 4. Sensitivity analyses of the conditional SHBG–HbA1c association.
(Points and horizontal bars show estimates and corresponding 95% confidence intervals from the primary and sensitivity models.)

## 3.5 SHBG within the Multidimensional Biomarker Space

In the 899-participant multiblock sample, SHBG correlated positively with HDL (ρ≈0.437) and negatively with GGT (ρ≈-0.316), uric acid (ρ≈-0.310), ALP (ρ≈-0.292), and hsCRP (ρ≈-0.265). The leading MFA dimensions explained 13.45%, 9.36%, and 7.28% of total inertia (30.10% cumulatively). Dimension 1 was dominated by biochemical, inflammation/lipid, and hematologic blocks, whereas the hormone block contributed 79.91% to dimension 3.

Within the hormone block, SHBG had a near-zero loading on the first principal component (-0.012) and larger loadings on PC4 (0.696) and PC5 (0.515). Under parallel analysis only the first three components were retained, implying that a strict unsupervised reduction could discard the focal exposure's dominant local direction. This supports retaining prespecified biomarkers alongside latent reduction rather than treating dimension reduction as a complete replacement for single-marker analysis.

After SHBG was removed and MFA rebuilt on the remaining 39 biomarkers, the first three dimensions explained 30.35%. Conditioning on these axes left an SHBG coefficient of -0.1062 (P=0.00342), approximately 73% of the magnitude of the primary coefficient. LASSO selected SHBG in 98.8% of 500 bootstrap resamples at λ.min, but only 38.8% at λ.1se. sPLS concentrated supervised information in its first component but retained all 40 biomarkers, indicating limited sparsity.

### 3.6 Internal Prediction and Latent Heterogeneity

Table 2. Performance of the three types of models under nested cross-validation (N=776)

| Model | MSE | RMSE | MAE | OOF $R^2$ |
|---|---|---|---|---|
| Clinical covariates | 0.492 | 0.7014 | 0.3767 | 0.1298 |
| 40 biomarkers | 0.445 | 0.6668 | 0.3780 | 0.2136 |
| Combined | 0.430 | 0.6559 | 0.3702 | 0.2392 |

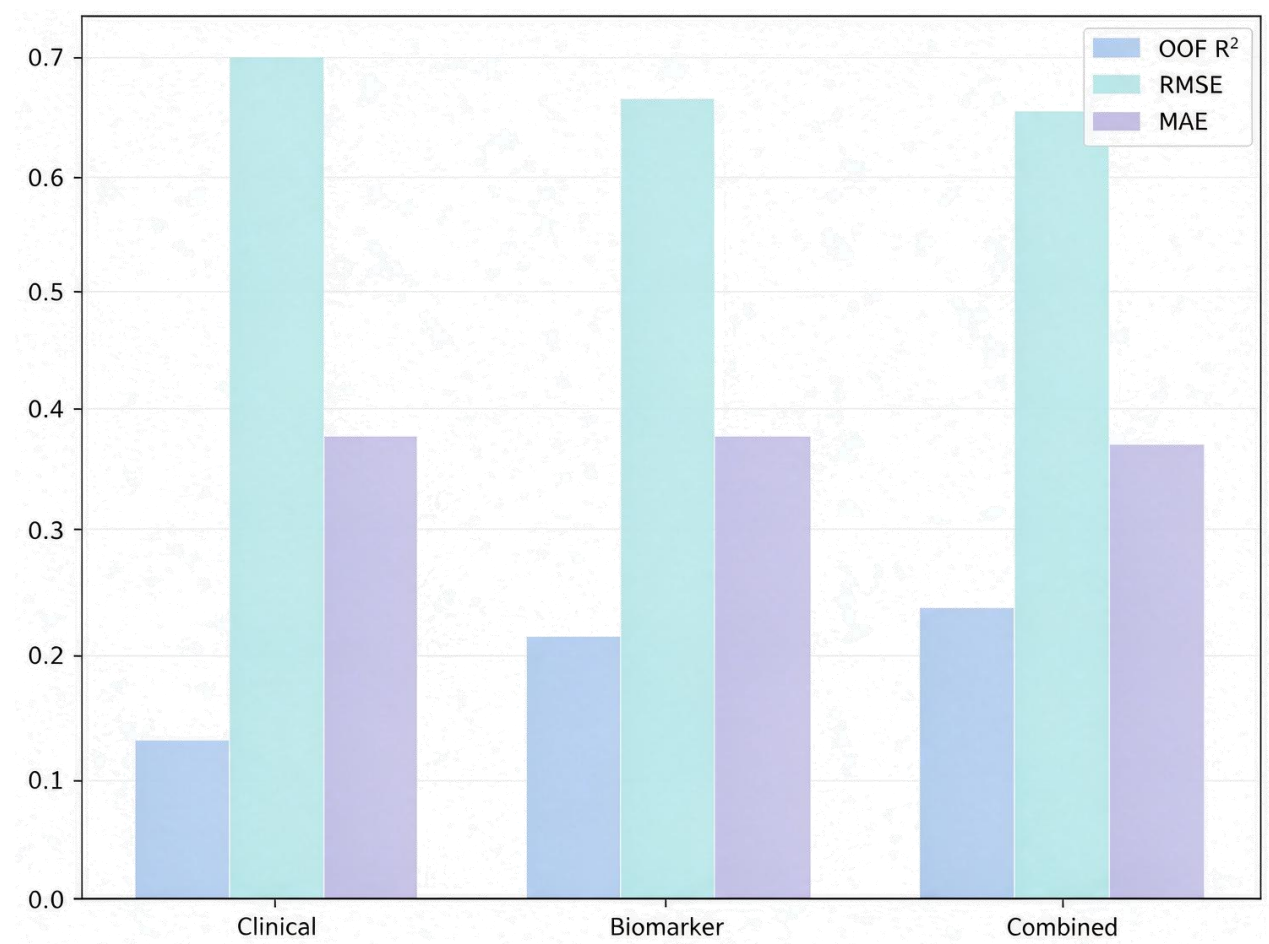

Figure 5. Nested cross-validation performance for clinical, biomarker-only, and combined models.

The combined model improved OOF $R^2$ by 0.1093 compared with the clinical model. The same direction persisted when self-reported diabetes was excluded and when HbA1c was restricted to < 6.5%, supporting an internal predictive increment that was not entirely driven by the small number of participants with marked hyperglycemia. No external validation was performed.

Outcome-guided sJIVE selected joint rank 0, with stable repeated cross-validation results, indicating that HbA1c-related information was concentrated in block-specific rather than cross-block joint components. FMR and MoE models converged, but all minimum class proportions were below 10%. Strict five-fold OOF $R^2$ was 0.1408 for the single quadratic guard versus 0.0984—0.1045 for the mixture models. Neural-network MoE fits did not converge and therefore were not interpreted.

## 4. Discussion

This study provides three main findings. First, in reproductive-age women, SHBG was inversely associated with continuous HbA1c after adjustment for major demographic, socioeconomic, and adiposity variables. Second, the association was nonlinear and steeper at low SHBG levels, while the linear slope became more negative with increasing age. Third, SHBG occupied a cross-domain endocrine–metabolic position within the 40-biomarker space and retained conditional information after the leading multiblock axes were controlled; the biomarker panel also improved internal predictive performance beyond clinical covariates.

The direction of the primary association is consistent with the broad female literature. Prospective evidence from Ding et al. showed lower future diabetes risk with higher SHBG. Hedderson et al. demonstrated that within-person SHBG trajectories across the menopause transition were associated with incident diabetes [8]. In younger women, CARDIA data linked higher SHBG with lower long-term metabolic syndrome risk [9]. The present study complements these findings by focusing on a younger, reproductive-age population and a continuous glycemic outcome rather than an incident disease endpoint.

The nonlinear pattern is potentially important. A single linear coefficient implies a constant change in HbA1c per relative increase in SHBG, whereas the present data suggest that the association is concentrated toward lower SHBG levels and attenuates at higher levels. This pattern is compatible with a biological or metabolic context in which the range of low SHBG identifies a more strongly adverse metabolic state, but the cross-sectional design prevents interpretation of the curve as a causal dose–response function.

The attenuation after adiposity and HOMA-IR adjustment supports substantial metabolic overlap. Stangl et al. reported that the association between low SHBG and type 2 diabetes is linked to visceral and liver fat using observational and Mendelian-randomization analyses [19]. The present attenuation with BMI, waist circumference, and HOMA-IR is therefore biologically coherent. The divergence from Crawford et al., who reported no straightforward SHBG–HOMA-IR relationship in 319 healthy cycling premenopausal women [11], may reflect differences in age composition, adiposity, laboratory measurement, and study selection.

The multiblock results add a different dimension. SHBG was not simply a representative of the dominant hormone axis: it had weak loading on the first hormone component, stronger loadings on later components, and stronger marginal connections to HDL, liver enzymes, uric acid, and inflammation. The persistence of the SHBG coefficient after removal of SHBG from MFA indicates that its association with HbA1c cannot be fully represented by the leading latent structure available in this biomarker panel. This result is consistent with recent mechanistic reviews emphasizing both sex-steroid-dependent and partly SHBG-specific pathways [2,5].

The prediction analysis should be interpreted cautiously. Nested cross-validation protects against optimistic performance caused by tuning on the same observations used for evaluation [17], and the combined model's OOF $R^2$ exceeded the clinical model by 0.1093. However, OOF performance within one NHANES cycle is not external validation and does not establish clinical utility. The mixture-model analyses reinforce a similar principle: lower BIC alone did not establish latent subgroups, because the fitted classes were small, failed the prespecified size criterion, and performed worse than the continuous quadratic guard out of sample.

## 5 Strengths and Limitations

Strengths include the use of the cycle-specific NHANES phlebotomy weights and complex survey variance structure; a focused reproductive-age female population; continuous modeling of both SHBG and HbA1c; complementary nonlinear analyses; and a multiblock workflow spanning structure, supervised selection, stability, and leakage-aware internal prediction. The latent-class analysis was deliberately guarded against overinterpretation by requiring minimum class size and strict OOF improvement over a continuous nonlinear model.

Limitations include the cross-sectional design; single measurements of SHBG and other hormones; incomplete characterization of menstrual phase, hormonal contraception, diet, medication use, and liver fat; a smaller fasting subsample; exploratory rather than fully design-based multiblock inference; and lack of external validation. The latent-class negative result indicates insufficient support in this sample, not proof that biologic heterogeneity does not exist.

## 6 Conclusions

Among nonpregnant women aged 20–44 years in NHANES 2021–2023, higher serum SHBG was associated with lower continuous HbA1c after adjustment for age, race/ethnicity, PIR, and BMI. The association was nonlinear, with the strongest inverse slope at low SHBG levels and attenuation at higher levels; the linear slope varied with age, but the evidence did not establish age modification of the entire nonlinear curve. Adiposity and insulin-resistance-related variables substantially overlapped with the association without supporting causal decomposition.

Within a 40-biomarker multiblock space, SHBG occupied an endocrine−metabolic intersection and retained conditional information beyond leading latent axes. The combined biomarker model demonstrated an internal predictive increment over clinical covariates, whereas sJIVE and mixture-model results did not support stable discrete biological subgroups. Longitudinal studies with repeated reproductive-endocrine and metabolic measurements, adipose and liver-fat phenotyping, and external validation are needed to determine temporality, mechanisms, and transportability.

## Data Availability

NHANES 2021—2023 public-use files are freely available from the National Center for Health Statistics.